\documentclass[11pt]{article}
\usepackage[margin=1in]{geometry}
\usepackage{amsmath,amsthm,amssymb}
\usepackage{graphicx}
\usepackage{wasysym}
\usepackage[linesnumbered,ruled,vlined]{algorithm2e}
\usepackage{hyperref}

{\bfseries}{\rmfamily}
\newtheorem{theorem}{Theorem}[section]
\newtheorem{definition}[theorem]{Definition}

\newtheorem{observation}[theorem]{Observation}
\newtheorem{corollary}{Corollary}[theorem]
\newtheorem{lemma}[theorem]{Lemma}
\newtheorem{proposition}[theorem]{Proposition}
\newcommand{\floor}[1]{\left\lfloor #1 \right\rfloor}
\newcommand{\ceil}[1]{\left\lceil #1 \right\rceil}

\usepackage{xcolor}
\definecolor{light-gray}{gray}{0.95}

\usepackage{array}
\newcolumntype{L}{>{$}l<{$}}

\DeclareMathOperator*{\argmax}{arg\,max}

\global\long\def\set{\operatorname{set}}
\global\long\def\st{\operatorname{st}}

\global\long\def\safe{\operatorname{safe}}
\global\long\def\unanimity{\operatorname{unanimity}}

\newcommand{\bfR}{\mathbf{R}}
\newcommand{\bfS}{\mathbf{S}}
\newcommand{\R}{\mathbb{R}}

\newcommand{\diam}{\mathrm{diam}}

\newcommand{\rank}{\mathrm{rank}}

\newcommand{\sync}{\mathtt{tsync}}
\newcommand{\SYNC}{\ensuremath{\textsc{sync}}}
\newcommand{\ASYNC}{\ensuremath{\textsc{async}}}
\newcommand{\nbar}{\overline{n}}
\newcommand{\calC}{\mathcal{C}}

\usepackage{subfiles}

\begin{document}

\title{Distributed Agreement in the Arrovian Framework}



\author{
Kenan Wood\thanks{Corresponding author}\\
\texttt{kewood@davidson.edu}\\
Davidson College
\and
Hammurabi Mendes\\
\texttt{hamendes@davidson.edu}\\
Davidson College
\and
Jonad Pulaj\\
\texttt{jopulaj@davidson.edu}\\
Davidson College
}

\date{}



\maketitle

\begin{abstract}
    Preference aggregation is a fundamental problem in voting theory, in which public input rankings of a set of alternatives (called \emph{preferences}) must be aggregated into a single preference that satisfies certain soundness properties. The celebrated Arrow Impossibility Theorem is equivalent to a distributed task in a synchronous fault-free system that satisfies properties such as respecting unanimous preferences, maintaining independence of irrelevant alternatives (IIA), and non-dictatorship, along with \emph{consensus} since only one preference can be decided.

    In this work, we study a weaker distributed task in which crash faults are introduced, IIA is not required, and the consensus property is relaxed to either $k$-set agreement or $\epsilon$-approximate agreement using any metric on the set of preferences. In particular, we prove several novel impossibility results for both of these tasks in both synchronous and asynchronous distributed systems. We additionally show that the impossibility for our $\epsilon$-approximate agreement task using the \emph{Kendall tau} or \emph{Spearman footrule} metrics holds under extremely weak assumptions.
\end{abstract}

\section{Introduction}\label{Sec:Intro}
Preference aggregation is a classical problem in voting theory where every voter publishes an input \emph{vote} (e.g. a single most-favorable candidate, a linear ranking of all the candidates, weighted rankings of candidates, etc.) and a typically centralized system aggregates the votes into a single decision outcome, according to some choice rule. Various soundness properties of preference aggregation algorithms are desirable, such as preservation of unanimous votes (unanimity), the notion that the outcome is not totally dictated by a small set of voters (non-dictatorship), and many more \cite{campbell-chapter1}. 

One of the first and most-celebrated fundamental results in voting theory was proved by Kenneth Arrow in 1951 \cite{arrow1951}. His theorem considers the preference aggregation problem where input votes and the decision outcomes consist of weak linear rankings of the \emph{alternatives} (candidates). In addition to unanimity and non-dictatorship, Arrow considered \emph{independence of irrelevant alternatives} (IIA), which requires that the outcome of the preference aggregation rule with respect to the relative ordering of any two alternatives should only depend on the voters' initial relative ordering of those two alternatives, and none of the other ``irrelevant'' pairs. Arrow's Impossibility Theorem asserts that no deterministic preference aggregation algorithm can satisfy unanimity, non-dictatorship, and IIA simultaneously. 

This result has received significant attention from the voting theory community. There are many generalizations of Arrow's theorem and a plethora of different proofs, including elementary proofs that directly exploit the unanimity and IIA axioms \cite{three-brief-proofs}, as well as others using more advanced mathematics like algebraic topology \cite{algebraic-topology-arrow-proof}, and fixpoint methods in metric space topology \cite{fixed-point-proof1, fixed-point-proof2}. 

There are recent advancements stemming from the methods used in distributed computing that yield better geometric/topological understanding of Arrow's Impossibility Theorem. Specifically, Lara, Rajsbaum, and Raventós-Pujo \cite{sergio2022, sergio2024} use techniques from combinatorial topology to prove a generalization of Arrow's theorem that sheds light on the result's geometric and topological structure. In general, combinatorial topology has also been successful in proving a myriad of impossibility and complete characterization results in distributed computing \cite{dist-comp-comb-topol, mendes-async-computability,sergio-continuous-tasks,mendes-phd}.

In this work, we introduce novel distributed tasks that combine aspects from well-studied problems such as \emph{set agreement} \cite{chaudhuri-k-set-init, mendes-k-set-agreement, sergio-k-set-failure-detectors, sergio-obstruction-free-k-set, sergio-opinion-number-set-agreement, k-set-tight-bounds} and \emph{approximate agreement} \cite{mendes-approx-agreement-DISC, mendes-approx-agreement-journal, sergio-topological-approx-agree, byzantine-approx-agree-graphs, wait-free-approx-agree-graphs, approx-agreement-simplicial-complexes}, as well as voting theoretic properties in the Arrovian framework.
Specifically, we study two tasks that require unanimity \emph{on the correct processes} along with either $k$-set agreement---meaning at most $k$ different preferences are decided---or $\epsilon$-approximate agreement---meaning any two decisions are within a distance $\epsilon$ apart, with respect to a specified metric. It is important to observe that the tasks presented here are significantly weaker than a na\"ive translation of the properties in Arrow's Impossibility Theorem; both the IIA and non-dictatorship properties are absent from our problem definition, and consensus is weakened to set or approximate agreement. In fact, we show in Section \ref{Sec:Dist-Agree-Pref-Agg} that a natural formulation of generalized non-dictatorship properties is actually implied by unanimity on the correct processes.

The main contributions of this paper are the following.
\begin{enumerate}
    \item In our main result, Theorem \ref{Thm:main-impossibility}, we formally show that under the relatively weak property that there exists some set of input preferences with a variably fine-grained cyclic structure (see Definitions \ref{Def:k-cyclic-pref-list} and \ref{Def:k-cyclic-profile}), neither of our Arrovian tasks are solvable for reasonable agreement parameters and sufficiently many alternatives, despite the apparent weakening of consensus.
    
    \item We also prove that for the special cases of $\epsilon$-approximate preference aggregation, when the distances between preferences are measured using the well-established Kendall tau \cite{kendall-tau1, kendall-tau2} or Spearman footrule \cite{footrule} metrics, no algorithm exists if $\epsilon$ is less than a certain large quantity, expressed in terms of the metric diameter, presented in Definition \ref{Def:diam}.
    

    \item We present a unified analysis that captures synchronous and asynchronous systems simultaneously, meant intentionally to shed light on the fundamental structural properties of the problem---such as cyclic preference patterns---that cause the impossibility.
\end{enumerate}

The rest of this paper is organized as follows. Section \ref{Sec:Background} provides the necessary background on relations (Subsection \ref{Subsec:relations}) and introduces \emph{distributed aggregation maps} in the context of synchronous fault-free distributed algorithms (Subsection \ref{Subsec:dist-agg-maps}). In Section \ref{Sec:Init-Observations}, we prove some initial observations which are simple generalizations of Arrow's theorem in the context of distributed aggregation maps. Section \ref{Sec:Dist-Agree-Pref-Agg} formally introduces the intersection of the Arrovian framework and set and approximate agreement tasks discussed above, and proves our remark that a non-dictatorship property is nonrestrictive.
The main results of this paper are in Section \ref{Sec:Main-Results}, and specifically, can be found in Theorem \ref{Thm:main-impossibility} and its corollaries.

\section{Background}\label{Sec:Background}
In this section, we present the necessary background on relations, and introduce the notation of a \emph{distributed} aggregation map and some of its relevant properties. Throughout this paper, we use the following notation: for a positive integer $n$, let $[n] \triangleq \{1, \dots, n\}$.

\subsection{Relations}\label{Subsec:relations}
A \emph{relation} on a set $X$ is a subset of $X \times X$. If $R$ is a relation we usually write $x \mathrel{R} y$ as shorthand for $(x,y) \in R$. A relation $R \subseteq X \times X$ is \emph{reflexive} provided $x \mathrel{R} x$ for all $x \in X$; it is \emph{antisymmetric} or \emph{strict} provided that $x \mathrel{R} y$ and $y \mathrel{R} x$ imply $x = y$ for all $x,y \in X$; $R$ is \emph{complete} if for every pair $x,y \in X$, it follows that $x \mathrel{R} y$ or $y \mathrel{R} x$;
$R$ is said to be \emph{transitive} provided that for every $x,y,z \in X$, $x \mathrel{R} y$ and $y \mathrel{R} z$ imply $x \mathrel{R} z$. 
Given a relation $R$ and elements $a,b$, we often write $a \succsim_{R} b$ for $a \mathrel{R} b$ and $a \succ_R b$ for $a \mathrel{R} b$ but $a \not \mathrel{R} b$.
The \emph{strict part of $R$} is defined by 
\[
\st(R) \triangleq \{(a,b) \in R: b \not\mathrel{R} a\},
\]
that is, the pairs in $R$ without equivalence.
So, $a \succ_{\st(R)} b$ if and only if $a \succ_{R} b$. If $R$ is any relation on $X$ and $Y \subseteq X$, we define the \emph{restriction of $R$ to $Y$} to be the subrelation $R|_{Y} \triangleq R \cap (Y \times Y)$ of $R$. Similarly, if $\mathbf{R} = (R_1, \dots, R_n)$ is a vector of relations on $X$ and $Y \subseteq X$, we define the restriction $\mathbf{R}|_{Y} \triangleq (R_1|_{Y}, \dots, R_n|_{Y})$.

The set of all reflexive complete transitive relations, called \emph{preferences}, on a set $X$ is denoted $P(X)$; the set of all antisymmetric reflexive complete transitive relations, called \emph{strict preferences}, on $X$ is denoted $L(X)$. A subset of $P(X)$ is called a \emph{domain}. A \emph{profile} on $X$ is an element of $\bigcup_{k=1}^\infty P(X)^k$; usually, we consider only profiles in $P(X)^n$ or $P(X)^{n-t}$, where $n$ is the number of processes and $t$ is the maximum number of faulty processes.

\subsection{Distributed Aggregation Maps}\label{Subsec:dist-agg-maps}
Consider a synchronous distributed message-passing system with $n$ processes $\{p_1, \dots, p_n\}$.
In this subsection, we consider only deterministic fault-free models of computation. 
In particular, we are interested in algorithms of the following type. Let $X$ be a finite set of \emph{alternatives} (also called candidates) and let $W_I, W_O \subseteq P(X)$ be domains of preferences; we denote the size of $X$ by $m$. Suppose processes take inputs from $W_I$, and after communicating, decide a preference in $W_O$ satisfying certain desirable properties of a distributed voting system.
Given these assumptions, distributed algorithms where processes have inputs in $W_I$ and decide values in $W_O$ are completely characterized by functions $F: W_I^n \to W_O^n$. 
To see this, since such algorithms are fault-free, every process has complete information after only one round of communication, so that the determinism of the algorithm guarantees the given map; the opposite direction is similarly trivial. This motivates the following definition.
\begin{definition}[Distributed Aggregation Map]\label{Def:dist-agg-map}
    A \emph{distributed aggregation map} on an $n$-process system with input domain $W_I \subseteq P(X)$ and output domain $W_O \subseteq P(X)$ is a map from $W_I^n$ to $W_O^n$.
\end{definition}

The properties we are interested in obtaining are characterized in the following definitions. For the rest of this section, let $F$ be an $n$-process distributed aggregation map with input domain $W_I$ and output domain $W_O$.
\begin{definition}[Unanimity]\label{Def:unanimity}
    We say that $F$ satisfies \emph{unanimity} if the following holds: for all $a,b \in X$ and for all $\bfR \in W_I^n$ such that $a \succ_{R_i} b$ for all $i \in [n]$, we have $a \succ_{F(\bfR)_i} b$ for all $i \in [n]$.
\end{definition}
\begin{definition}[IIA]\label{Def:IIA}
    The map $F$ satisfies \emph{independence of irrelevant alternatives} (IIA) if for all $a,b \in X$ and all $\bfR, \bfS \in W_I^n$ such that $\bfR|_{\{a,b\}} = \bfS|_{\{a,b\}}$, it follows that $F(\bfR)|_{\{a,b\}} = F(\bfS)|_{\{a,b\}}$.
\end{definition}

The following definition of decisiveness is standard terminology in the voting theory community \cite{campbell-chapter1}, but the motivation is as follows: a set $S$ is \emph{decisive} if all of the strict rankings between pairs are always dictated by the strict relations of the inputs in $S$.
\begin{definition}[Decisive]\label{Def:decisive}
    A set $S \subseteq [n]$ is said to be \emph{decisive} if $S$ is nonempty and the following holds: for all $a,b \in X$ and all $\bfR \in W_I^n$ such that $a \succ_{\bfR_i} b$ for all $i \in S$, it follows that $a \succ_{F(\bfR)_j} b$ for all $j \in [n]$.
\end{definition}\label{Def:dictatorship}
\begin{definition}[Dictatorship]\label{def-decisive}
    If $F$ has a decisive set of size at most $k$, we say that $F$ is a \emph{$k$-dictatorship} or that $F$ is {$k$-dictatorial}; if $k=1$, we say that $F$ is simply a \emph{dictatorship} or that $F$ is \emph{dictatorial}.
\end{definition}

In this paper, we are interested in set agreement and approximate agreement relaxations of the consensus property that is usually assumed in the Arrovian framework.
Given a vector or list $v$, define $\set(v)$ to be the set of entries in $v$.
\begin{definition}[Set Agreement]\label{Def:set-agree}
    If for all $\bfR \in W_I^n$, we have $|\set(F(\bfR))| \le k$, then we say that $F$ satisfies \emph{$k$-set agreement}. The property of $1$-set agreement is called \emph{consensus}.
\end{definition}

Before describing the approximate agreement condition, we introduce some definitions and corresponding notation commonly used throughout this paper.

\begin{definition}[Metric Diameter]\label{Def:diam}
    If $d$ is a metric on a finite set $Y$ and $A \subseteq Y$, we define the \emph{diameter} of $A$ with respect to $d$ by $\diam_d(A) \triangleq \max_{x,y \in A} d(x, y)$. The diameter of a list $v$ of elements of $Y$ is defined to be the diameter of the set of values it contains, and is written $\diam_d(v)$.
\end{definition}
\begin{definition}[Approximate Agreement]\label{Def:approx-agree}
    If $d$ is a metric on a domain containing $W_O$, then we say that $F$ satisfies \emph{$\epsilon$-agreement} (for $\epsilon \ge 0$) with respect to $d$ if every $\bfR \in W_I^n$ satisfies $\diam_d(F(\bfR)) \le \epsilon$.
\end{definition}

Our main results in Section \ref{Sec:Main-Results} pay special attention to the following natural metrics~\cite{kendall-tau1, kendall-tau2, footrule} on $L(X)$ which are useful measures of distance in the context of approximate agreement. 
The Kendall tau metric measures the number of \emph{pairs} of alternatives that differ, while Spearman's footrule measures that \emph{cumulative} distance between the ranks of each of the alternatives.
These metrics are formally described below.
\begin{definition}[Kendall tau]\label{Def:kendall-tau}
    Define the \emph{Kendall tau} metric on $L(X)$, denoted $\mathtt{KT}$, by 
    $\mathtt{KT}(R, S) \triangleq |\{(a,b) \in X \times X: a \succ_R b \wedge b \succ_S a\}|$
    for all $R, S \in L(X)$.
\end{definition}

\begin{definition}[Rank and Spearman's footrule]\label{Def:spearman-footrule}
    Given $R \in L(X)$ and $a \in X$, define the \emph{rank} of $a$ in $R$ by $\rank_R(a) \triangleq |\{b \in X: b \succsim_R a\}|$. The \emph{Spearman footrule} on $L(X)$ is a metric $\mathtt{SF}$ defined by $\mathtt{SF}(R, S) \triangleq \sum_{a \in X} |\rank_R(a) - \rank_S(a)|$ for all $R,S \in L(X)$.
\end{definition}

The following results with respect to diameter in the Kendall tau or Spearman footrule metrics can be found in \cite{footrule}, but are otherwise easy to prove.
\begin{proposition}\label{Prop:kendall-spearman-diam}
    If $|X| = m$, then $\diam_{\mathtt{KT}}(L(X)) = \frac{m^2-m}{2}$ and $\diam_{\mathtt{SF}}(L(X)) = \floor{\frac{m^2}{2}}$.
\end{proposition}

Since we are interested in distributed analogues to Arrow's Theorem, we introduce the following definition.
\begin{definition}\label{Def:AC-domain}
    We say that a domain $W \subseteq P(X)$ is \emph{Arrow-Complete} (AC) if every distributed aggregation map on $n$ processes with input domain $W$ and output domain $P(X)$ that satisfies unanimity, IIA, and consensus is dictatorial.
\end{definition}

Arrow's Impossibility Theorem then states the following.
\begin{theorem}[Arrow]\label{Thm:Arrow}
    If $n \ge 2$ and $m \ge 3$, then $P(X)$ is AC.
\end{theorem}
In \cite{campbell-chapter1}, Arrow's theorem has also been generalized to show that any domain satisfying a certain \emph{chain rule} is AC if $n \ge 2$ and $m \ge 3$.

\section{Initial Observations}\label{Sec:Init-Observations}

In this section, we state and prove our initial observations as a starting point for discussing our main theorem and corollaries in Section \ref{Sec:Main-Results}. In particular, we prove two impossibility results related to distributed aggregation maps when the consensus condition is weakened to either set agreement or approximate agreement. The core of both arguments relies on a simple coordinate-by-coordinate reduction of a distributed aggregation map. Although we find these preliminary results interesting, the main contributions of this paper can be found at the end of Section \ref{Sec:Main-Results} (see Theorem \ref{Thm:main-impossibility} and its corollaries in Section \ref{Sec:Main-Results}). In particular, our main results are impossibility theorems in fault-prone distributed systems with either synchronous or asynchronous communication between processes. 

\begin{proposition}\label{Prop:k-set-perfect-sync}
    Let $k$ be a positive integer such that $k < n$ and $k < |W|$, where $W$ is an AC domain. Furthermore, assume there is a set $P \subseteq W$ of size at least $k+1$ such that for any two distinct $R, R' \in P$, there exists $a,b \in X$ satisfying $a \succ_{R} b$ and $b \succ_{R'} a$. Every distributed aggregation map on $W$ satisfying $k$-set agreement, unanimity, and IIA is $k$-dictatorial.
\end{proposition}

\begin{proposition}\label{Prop:approx-agree-perfect-sync}
    Let $W$ be an AC domain that contains some strict preference. Let $d$ be a metric on $W$. If $\epsilon < \diam_d(W \cap L(X))$, then every distributed aggregation map on $W$ satisfying $\epsilon$-agreement (with respect to $d$), unanimity, and IIA is dictatorial.
\end{proposition}

The key observation underlying the proofs of both of these results is the following lemma. The proof is based on a simple coordinate-by-coordinate reduction from any distributed aggregation map to a distributed aggregation map satisfying consensus. In observing that this reduction preserves unanimity and IIA, we exploit Arrow-Completeness to show that each reduced map is a dictatorship. This shows that every output coordinate is in a sense ``dictated'' by an input coordinate, which is unique under a very weak condition. It is important to note that we are not claiming to reprove Arrow's theorem in any way, and instead, we are building on a domain that already satisfies Arrow's theorem.
\begin{lemma}\label{Lem:partial-dictator}
    Suppose $W$ is an AC domain. Let $F: W^n \to P(X)^n$ be a distributed aggregation map. Suppose $F$ satisfies unanimity and IIA, and let $j \in [n]$. Then we have the following.
    \begin{enumerate}
        \item Then there exists some $i \in [n]$ such that for all $a,b \in X$ and $\bfR \in W^n$ such that $a \succ_{R_i} b$, we have $a \succ_{F(\bfR)_j} b$.
        \item If $W$ contains some two preferences $R, R'$ and $a,b \in X$ such that $a \succ_{R} b$ and $b \succ_{R'} a$, then the $i$ above is unique.
    \end{enumerate}
\end{lemma}
\begin{proof}
    Construct a map $F^j: W^n \to P(X)^n$ by setting $F^j(\bfR) = (F(\bfR)_j)_{i \in [n]}$ for all $\bfR \in W^n$. It is clear $F^j$ satisfies consensus. It is also easy to show that since $F$ satisfies unanimity and IIA, $F^j$ also satisfies unanimity and IIA. Since $W$ is AC, it follows that $F^j$ is dictatorial. Thus (1) follows from the construction of $F^j$ and Definition~\ref{Def:decisive}.

    To show the second part, suppose $i$ and $i'$ be two elements of $[n]$ satisfying the above criteria. Consider a preference profile $\bfR$ such that $R_i = R$ and $R_{i'} = R'$, where $R$ and $R'$ are defined in the lemma statement. Let $a,b \in X$ such that $a \succ_{R} b$ and $b \succ_{R'} a$. By the assumption on $i$ and $i'$, we know that $a \succ_{F(\bfR)_j} b$ (as $R_i = R$ and $i$ satisfies (1)) and $b \succ_{F(\bfR)_j} a$ (as $R_{i'} = R'$ and $i'$ satisfies (1)), which is a contradiction. Thus (2) follows.
\end{proof}

We are now ready to prove Proposition \ref{Prop:k-set-perfect-sync} and Proposition \ref{Prop:approx-agree-perfect-sync}.
\begin{proof}
    [Proof of Proposition \ref{Prop:k-set-perfect-sync}] Let $P$ be defined as in the theorem statement. By Lemma~\ref{Lem:partial-dictator}, every $j \in [n]$ can be uniquely mapped to some $\delta(j) \in [n]$ such that for all $a,b \in X$ and $\bfR \in W^n$ such that $a \succ_{R_{\delta(j)}} b$, we have $a \succ_{F(\bfR)_j} b$. Let $S = \{\delta(j): j \in [n]\}$. It is clear by construction that $S$ is a decisive set for $F$, so it remains to show that $|S| \le k$. Suppose for contradiction that $|S| \ge k+1$. Let $S' \subseteq S$ such that $|S'| = k+1$. It follows that there exists an injection $g: S' \to P$. Also, fix an injection $\Delta: S' \to [n]$ such that $\Delta(i) \in \delta^{-1}(i)$ for all $i \in [n]$. Construct any profile $\bfR \in W^n$ such that for all $i \in S'$, we have $R_i = g(i) \in P$. Suppose $i, i' \in S'$ are distinct. Consider $\Delta(i)$ and $\Delta(i')$, which are distinct as $\Delta$ is injective. Observe that since $g$ is a bijection, $g(i) \ne g(i')$. By definition of $P$ and noting that $g(i), g(i') \in P$, there exists $a,b \in X$ such that $a \succ_{R_i} b$ and $b \succ_{R_{i'}} a$. As $\delta(\Delta(i)) = i$ and $\delta(\Delta(i')) = i'$, the definition of $\delta$ implies that $a \succ_{F(\bfR)_{\Delta(i)}} b$ and $b \succ_{F(\bfR)_{\Delta(i')}} a$. This shows that $F(\bfR)_{\Delta(i)} \ne F(\bfR)_{\Delta(i')}$; that is, every $F(\bfR)_{\Delta(i)}$ is distinct over all $i \in S'$. In particular, this shows that
    \[
    |\set(F(\bfR))| \ge |\{F(\bfR)_{\Delta(i)}: i \in S'\}| = |S'| = k+1 > k.
    \]
    This contradicts the $k$-set agreement property of $F$. Hence $|S| \le k$, which shows $F$ is $k$-dictatorial.
\end{proof}

\begin{proof}
    [Proof of Proposition~\ref{Prop:approx-agree-perfect-sync}] Suppose $F$ is a distributed aggregation map on $W$ that satisfies unanimity and IIA. Suppose for contradiction that $F$ is not dictatorial. For each $j \in [n]$, let $\delta(j) \in [n]$ such that for all $a,b \in X$ and $\bfR \in W^n$ such that $a \succ_{R_{\delta(j)}} b$, we have $a \succ_{F(\bfR)_j} b$, which is well-defined by Lemma \ref{Lem:partial-dictator}. Since $F$ is not dictatorial, not all values of $\delta(j)$ for $j \in [n]$ are equal, so that there exists distinct $j,j' \in [n]$ where $\delta(j) \ne \delta(j')$. Let $i = \delta(j)$ and $i' = \delta(j')$. This implies that for all $\bfR \in W^n$ if $R_i \in L(X)$, then $F(\bfR)_j = R_i$, and if $R_{i'} \in L(X)$, then $F(\bfR)_{j'} = R_{i'}$. 
    
    So let $R, R' \in W\cap L(X)$ such that $d(R, R') = \mathrm{diam}(W \cap L(X))$.
    Construct a profile $\bfR \in W^n$ such that $R_i = R$ and $R_{i'} = R'$. As $R_i \in L(X)$, the above remark shows that $F(\bfR)_j = R$. Similarly, $F(\bfR)_{j'} = R'$. By construction of $R$ and $R'$, if $\epsilon < \mathrm{diam}(W \cap L(X))$, then $F$ does not satisfy $\epsilon$-agreement with respect to $d$, a contradiction.
\end{proof}

\section{Distributed Set and Approximate Preference Aggregation}\label{Sec:Dist-Agree-Pref-Agg}

In this section and the next, we study distributed aggregation algorithms in the presence of process failures, in both synchronous and asynchronous communication models. The previous impossibility theorems in Section \ref{Sec:Init-Observations} were only in the synchronous fault-free case, so naively introducing failures into the system only makes the task at hand more difficult, and so the impossibility trivially holds. Hence we will discard the IIA property, as it is typically viewed as the most restrictive and least necessary for preference aggregation. Additionally, we discard the dictatorship properties as well, only requiring unanimity and agreement. We will find in this section and the next that we still obtain a plethora of impossibilities, despite this apparent simplification.

For the rest of this paper, consider a set of $n$ processes, $\{p_1, \dots, p_n\}$, at most $t$ (with $1 \le t < n$) of them suffering crash failures. Let $X$ be any finite set of $m \ge 2$ alternatives. Let $W_I, W_O \subseteq P(X)$ be input and output domains, respectively. The \emph{identity} of process $p_i$ is defined to be $i$. Let $C$ be the set of correct process identities in a given execution of a distributed algorithm. 

We focus on the following two problems, which consider distributed aggregation functions in a message-passing distributed system in the crash failure model.

\begin{definition}[$k$-Set Preference Aggregation]\label{Def:k-set-pref-problem}
    Let $k \ge 1$. The \emph{$k$-set preference aggregation} task with respect to $W_I, W_O$ has the following specifications. Each process $p_i$ selects a \emph{private} input preference $R_i \in W_I$. Every correct process $p_i$ decides a value $S_i \in W_O$ satisfying:
    \begin{itemize}
        \item $k$-set agreement. At most $k$ different orders are decided: $|S_i: i \in C| \le k$.
        \item Unanimity. For all $a,b \in X$, if every correct $p_i$ has $a \succ_{R_i} b$, then $a \succ_{S_i} b$ for all $i \in C$.
    \end{itemize}
\end{definition}

\begin{definition}[$\epsilon$-Approximate Preference Aggregation]\label{Def:approx-pref-problem}
    Let $\epsilon \ge 0$; let $d$ be a metric on a subset of $P(X)$ containing $W_O$. The \emph{$\epsilon$-approximate preference aggregation} task with respect to $W_I, W_O, d$ has the following specifications. Each process $p_i$ selects a \emph{private} input preference $R_i \in W_I$. Every correct process $p_i$ decides a value $S_i \in W_O$ satisfying:
    \begin{itemize}
        \item $\epsilon$-approximate agreement. All correct decisions are at most $\epsilon$ apart: $\diam_d(\{S_i: i \in C\}) \le \epsilon$.
        \item Unanimity. For all $a,b \in X$, if every correct $p_i$ has $a \succ_{R_i} b$, then $a \succ_{S_i} b$ for all $i \in C$.
    \end{itemize}
\end{definition}

Even without the IIA and non-dictatorship properties seen in Propositions \ref{Prop:k-set-perfect-sync} and \ref{Prop:approx-agree-perfect-sync} (the deterministic synchronous fault-free case), a significant amount of structure is still imposed on algorithms solving either of these tasks, particularly because of the strength of this unanimity property together with an agreement condition, as we shall see in Section \ref{Sec:Main-Results}.

Another reason for the lack of non-dictatorship criteria in $\epsilon$-approximate and $k$-set preference aggregation is that these properties are often implied by the unanimity condition. We make this precise in the following definition and proposition, noting that its proof is based on an indistinguishability argument commonly seen in distributed computing \cite{Fich2003, sergio-indistinguishability}.

A distributed algorithm $A$ is \emph{$k$-dictatorial} provided that the following holds for all admissible executions: there is a nonempty set $T \subseteq [n]$ with $|T| \le k$ such that
if every correct process $p_i$ has input $R_i$ and output $S_i$ and $T \subseteq C$, then $a \succ_{R_i} b$ for all $i \in T$ ($a,b \in X$) implies $a \succ_{S_i} b$ for all $i \in C$. 
A domain $W \subseteq P(X)$ is \emph{non-trivial} if there are two preferences $R, S \in W$ and two alternatives $a,b \in X$ such that $a \succ_R b$ and $b \succ_S a$.

\begin{proposition}\label{Prop:no-decisive-k-coalition}
    Suppose $W_I$ is non-trivial. Any algorithm in any synchrony model that satisfies unanimity is not $k$-dictatorial if $1 \le k \le t$.
\end{proposition}
\begin{proof}
    Suppose $A$ is an algorithm that satisfies unanimity, and assume $1 \le k \le t$.
    Let $R, R' \in W_I$ and $a,b \in X$ such that $a \succ_R b$ and $b \succ_{R'} a$. Let $T \subseteq [n]$ such that $1 \le |T| \le k$. Consider an execution $\Xi$ of $A$ where every process is non-faulty and every process $p_i$ for $i \in T$ has input $R$ and every other process has input $R'$. Since $|T| \le k \le t$, there exists an admissible execution $\Xi'$ that is identical to $\Xi$ except the set of faulty processes is precisely $T$. By unanimity, in $\Xi'$, processes $p_i$ with $i \in [n]\setminus T$ must decide $S_i' \in W_O$ satisfying $b \succ_{S_i'} a$. Since $\Xi$ and $\Xi'$ are indistinguishable executions for any $p_i$ with $i \in [n]\setminus T$, it follows that each such $p_i$ decides some $S_i \in W_O$ satisfying $b \succ_{S_i} a$. Since $a$ is ranked higher than $b$ for all inputs of processes with identity in $T$, this shows that $A$ is not $k$-dictatorial.
\end{proof}

We will make use of the following synchrony notation in the next sections. Define the \emph{synchrony} of a distributed system to be $\SYNC$ if the system is synchronous and $\ASYNC$ if the system is asynchronous. Let the synchrony of the distributed system at hand be denoted $\sync$. 
Additionally, we say that an execution of a distributed algorithm in a particular model of computation (synchrony and maximum number of faults) is \emph{admissible} if the execution satisfies the synchrony requirements and its number of faults is at most the maximum number of faults permissible by the model of computation.

\section{Arrovian Impossibilities in Synchronous and Asynchronous Systems}\label{Sec:Main-Results}
In this section, we present strong impossibility results for both synchronous and asynchronous systems and both $k$-set and $\epsilon$-approximate preference aggregation tasks. We begin with a simple definition that allows us to treat both synchrony models simultaneously.


\begin{definition}[Synchronous Process Number]\label{Def:sync-proc-num}
    Define the \emph{synchronous process number} $\overline{n}(\sync)$ of a synchrony $\sync \in \{\SYNC, \ASYNC\}$ by
    \[
    \overline{n}(\sync) \triangleq \begin{cases}
        n, & \text{if } \sync = \SYNC\\
        n-t, & \text{if } \sync = \ASYNC.
    \end{cases}
    \]
\end{definition}

Next we describe a convenient map that captures the relation between input and output of correct processes in an arbitrary admissible execution. Note that these reductions are not topological in nature as in \cite{dist-comp-comb-topol, k-set-tight-bounds, mendes-k-set-agreement, mendes-async-computability}, and are merely convenient tools used in our impossibility results. 

\begin{definition}[Execution Map: $\SYNC$]\label{Def:alg-reduction-SYNC}
    If $A$ is a distributed algorithm in the $\SYNC$ synchrony model with inputs in $W_I$ and outputs in $W_O$, define a map $\mathtt{F}^\SYNC_A: W_I^n \to W_O^n$ as follows: for each $\bfR \in W_I^n$, deterministically fix an execution of $A$ where all processes are correct and each $p_i$ has input $R_i$; let $S_i$ be the output of each $p_i$, and set $\mathtt{F}^\SYNC_A(\bfR) \triangleq \bfS = (S_1, \dots, S_n)$.
\end{definition}

In the following definition, we choose the set of $n-t$ correct processes in the given executions to be $p_1, \dots, p_{n-t}$ (and the faulty set to be $p_{n-t+1}, \dots, p_n$) for notational convenience, but this labeling is rather arbitrary.

\begin{definition}[Execution Map: $\ASYNC$]\label{Def:alg-reduction-ASYNC}
    Let $A$ be an algorithm in $\ASYNC$ synchrony model. Define a map $\mathtt{F}^\ASYNC_A : W_I^{n-t} \to W_O^{n-t}$ by setting $\mathtt{F}^\ASYNC_A(\bfR)$ for each $\bfR \in W_I^{n-t}$ as follows:
    deterministically fix an admissible execution of $A$ where $p_{n-t+1}, \dots, p_n$ are the $t$ faulty processes that crash before sending any messages, and $p_i$ has input $R_i$ for $i \in [n-t]$, and the correct processes $p_1, \dots, p_{n-t}$ communicate perfectly synchronously for the duration of the execution; let $S_i$ be the decided value of $p_i$; let $\mathtt{F}^\ASYNC_A(\bfR) \triangleq (S_1, \dots, S_{n-t}) = \bfS$.
\end{definition}

In either synchrony cases for $\sync \in \{\SYNC, \ASYNC\}$ of the above reductions, Definition \ref{Def:sync-proc-num} shows that the reduced map is from $W_I^{\nbar(\sync)}$ to $W_O^{\nbar(\sync)}$. Note that even when $A$ is a nondeterministic algorithm in the above definitions, a deterministic execution can still be chosen. For the rest of this section, fix a synchrony model $\sync \in \{\SYNC, \ASYNC\}$.
\begin{observation}\label{Lem:k-set-reduction}
    If $A$ is an algorithm that satisfies $k$-set agreement in the $\sync$ synchrony model, then the map $\mathtt{F}^\sync_A$ satisfies $k$-set agreement.
\end{observation}

\begin{observation}\label{Lem:approx-reduction}
    If $A$ is an algorithm that satisfies $\epsilon$-approximate agreement in the $\sync$ synchrony model, then the map $\mathtt{F}^\sync_A$ satisfies $\epsilon$-approximate agreement.
\end{observation}


The following definition of $u$-unanimity for distributed aggregation maps can be seen as a kind of unanimity that is preserved under $u$-of-$\nbar(\sync)$ thresholds.

\begin{definition}[$u$-Unanimity]
    A map $F: W_I^{\nbar(\sync)} \to W_O^{\nbar(\sync)}$ satisfies \emph{$u$-unanimity} for an integer $u$ if for all $a,b \in X$ and $\bfR \in W_I^{\nbar(\sync)}$, the set $T \triangleq \{i \in [\nbar(\sync)]: a \succ_{R_i} b\}$ satisfying $|T| \ge u$ implies that $a \succ_{F(\bfR)_i} b$ for all $i \in T$.
\end{definition}

The next auxiliary lemma is a simple application of a classical indistinguishability argument in distributed computing \cite{Fich2003, sergio-indistinguishability}, and connects distributed unanimity with $u$-unanimity for appropriate $u$.
\begin{lemma}\label{Lem:unanimity-reduction}
    If $A$ is an algorithm that satisfies unanimity in the $\sync$ synchrony model, then the map $\mathtt{F}^\sync_A$ satisfies $(\nbar(\sync) - t)$-unanimity.
\end{lemma}
\begin{proof}
    First, suppose $\sync = \ASYNC$.
    Consider any $a,b \in X$ and $\bfR \in W_I^{\nbar(\sync)} = W_I^{n-t}$. Suppose the set $T \triangleq \{i \in [n-t]: a \succ_{R_i} b\}$ satisfies $|T| \ge \nbar(\sync) - t = n-2t$. Let $\Xi$ be the execution of $A$ that defines $\mathtt{F}^\ASYNC_A(\bfR)$. Construct a new execution $\Xi'$ that sends and receives the same messages as $\Xi$ (and in the same order) but the set of faulty process identities is $[n-t] \setminus T$ instead of $\{n-t+1, \dots, n\}$, and the processes with identity greater than $n-t$ have their messages delayed until after every other process has decided; we may assume that each $p_i$ for $i > n-t$ has input $R_i$ satisfying $a \succ_{R_i} b$, so that every $i \in C$ satisfies $a \succ_{R_i} b$. Immediately, we know that processes $p_i$ for $i \in T$ must decide $\mathtt{F}_A^\ASYNC(\bfR)_i$ in $\Xi'$. Furthermore, there are $|[n-t]\setminus T| = (n-t) - |T| \le (n-t) - (n-2t) = t$ faulty processes in $\Xi'$ since $|T| \ge n-2t$. This implies that the execution $\Xi'$ is admissible in the $\ASYNC$ model since asynchronous communication delays may be unbounded (but still finite). It follows that every $i \in T$ satisfies $a \succ_{\mathtt{F}_A^\ASYNC(\bfR)_i} b$ since $p_i$ decides $\mathtt{F}_A^\ASYNC(\bfR)_i$ in $\Xi'$ and $A$ respects unanimity. Hence $\mathtt{F}_A^\sync(\bfR)$ satisfies $(\nbar(\sync)-t)$-unanimity.

    The proof for the case when $\sync = \SYNC$ is similar, but we include it here for completeness. Suppose $\sync = \SYNC$. Suppose $A$ is an algorithm with inputs from $W_I$ and outputs from $W_O$ that satisfies unanimity. Let $a,b \in X$ and $\bfR \in W_I^n$; let $T \triangleq \{i \in [n]: a \succ_{R_i} b\}$, and assume $|T| \ge n-t$. Let $\Xi$ be the execution of $A$ that defines $\mathtt{F}_A^\SYNC(\bfR)$ from Definition \ref{Def:alg-reduction-SYNC}. Let $\Xi'$ be an execution of $A$ obtained from $\Xi$ by letting each $p_j$ for $j \in [n]\setminus T$ be faulty but still send and receive exactly the same messages as in $\Xi$. Immediately by construction, for all $i \in T$, $p_i$ decides $\mathtt{F}^\SYNC_A(\bfR)_i$ in $\Xi'$. Since $|T| \ge n-t$, we know $|[n] \setminus T| \le t$, which shows that $\Xi'$ is an admissible execution of $A$. Since, in $\Xi'$, every correct process $p_i$ (for $i \in T$) has $a \succ_{R_i} b$, and since $A$ satisfies unanimity, it follows that every $p_i$ for $i \in T$ decides a preference that ranks $a$ above $b$. It follows that $a \succ_{\mathtt{F}^\SYNC_A(\bfR)_i} b$ for all $i \in T$, as desired. Hence $\mathtt{F}^\SYNC_A(\bfR)_i$ satisfies $(\nbar(\sync)-t)$-unanimity.
\end{proof}


A simple consequence of our auxiliary results thus far is the following. Suppose $t \ge n/2$ and let $A$ be an algorithm that satisfies unanimity. Then by Lemma \ref{Lem:unanimity-reduction}, $F = \mathtt{F}^{\ASYNC}_A$ satisfies $0$-unanimity, so that if $i \in [n-t]$ and $\bfR \in W_I^{n-t}$ has $R_i \in L(X)$, then $F(\bfR)_i = R_i$. Hence, $k$-set preference aggregation is impossible in the $\ASYNC$ synchrony model as long as $k < n-t$ and $|W_I \cap L(X)| \ge n-t$. Similarly, $\epsilon$-approximate preference aggregation is impossible in the $\ASYNC$ synchrony model if $W_I \cap L(X) \ne \emptyset$ and $\epsilon < \diam_d(W_I \cap L(X))$. Hence most of the interesting cases in the asynchronous communication model are when $t < n/2$.




Our main results in this section rely on the definitions below, inspired by the Mendes--Herlihy algorithm \cite{mendes-approx-agreement-DISC} for approximate agreement in $\R^d$, except with minor differences. 

\begin{definition}[Unanimity Set]\label{Def:unanimity-set}
    For an indexed set $M = \{R_\alpha\}_{\alpha\in J} \subseteq W_I$, define the \emph{unanimity set} of $M$ by 
    \[
    \unanimity(M) \triangleq \{S \in W_O: \forall a, b \in X, [(\forall \alpha\in J, a \succ_{R_\alpha} b) \implies a \succ_S b]\}.
    \]
\end{definition}

Thus, given an indexed set $M = \{R_\alpha\}_{\alpha \in J} \subseteq W_I$ for some $J \subseteq [n]$, the unanimity set of $M$ is the set of admissible output rankings required by unanimity, when process $p_\alpha$ has input $R_\alpha$. The next lemma introduces the concept of \emph{safe area} in this context.

\begin{definition}[Safe Area]\label{Def:safe-area}
    Let $M = \{R_\alpha\}_{\alpha \in J} \subseteq W_I$, where $J \subseteq [n]$ and $|J| > t$. For each $i \in J$, let
    \[
    \safe_i(M) \triangleq \bigcap_{\substack{T \subseteq J:\\ |T| = |J|-t\\ i \in T}} \unanimity(\{R_\alpha: \alpha \in T\}), 
    \]
    called the \emph{safe area} of $M$ with respect to $p_i$.
\end{definition}

We often slightly abuse notation by treating a tuple $(x_i)_{i=1}^k$ as an indexed set $\{x_i\}_{i \in [k]}$, as we do in the following lemma.
This next lemma shows the connection between unanimity of an algorithm and the safe area concept above.

\begin{lemma}\label{Lem:map-unanimity-to-safe-area}
    Let $F: W_I^{\nbar(\sync)} \to W_O^{\nbar(\sync)}$ be a distributed aggregation map that satisfies $(\nbar(\sync)-t)$-unanimity. Then for all $i \in [\nbar(\sync)]$ and all $\bfR \in W_I^{\nbar(\sync)}$, we have $F(\bfR)_i \in \safe_i(\bfR)$.
\end{lemma}
\begin{proof}
    Let $i \in [\nbar(\sync)]$ and $\bfR \in W_I^{\nbar(\sync)}$. Let us prove that $F(\bfR)_i \in \safe_i(\bfR)$. Suppose $T \subseteq [\nbar(\sync)]$ such that $i \in T$ and $|T| = \nbar(\sync) - t$. It suffices to show that $F(\bfR)_i \in \unanimity(\{R_\alpha: \alpha \in T\})$. To this end, suppose $a,b \in X$ such that for all $\alpha \in T$, $a \succ_{R_\alpha} b$. Since $F$ satisfies $(\nbar(\sync)-t)$-unanimity and $|T| \ge \nbar(\sync) - t$, it follows that $a \succ_{F(\bfR)_i} b$ as $i \in T$. This shows that $F(\bfR)_i \in \unanimity(\{R_\alpha: \alpha \in T\})$, as desired.
\end{proof}

To show an impossibility, we seek cases where $\safe_i(\bfR) = \{R_i\}$ for all $i$. This leads to the notion of a \emph{$k$-cyclic} profile, expressed through Definitions \ref{Def:k-cyclic-pref-list} and \ref{Def:k-cyclic-profile}. For convenience in the following definition and in the proof of Lemma \ref{Lem:cyclic-implies-safe} below, we use the following shorthand notation. If $X_1$ and $X_2$ are disjoint nonempty subsets of $X$ and $R \in W_I$, we write $X_1 \succ_{R} X_2$ as shorthand for $\forall x_1 \in X_1, \forall x_2 \in X_2, x_1 \succ_{R} x_2$; that is, we write $X_1 \succ_{R} X_2$ if and only if $R$ ranks every element of $X_1$ above every element of $X_2$. 

\begin{definition}[Cyclic Preference List]\label{Def:k-cyclic-pref-list}
    Let $1 \le k \le m$. A \emph{$k$-cyclic preference list} in $W_I$ is a list of $k$ distinct preferences $R_1, \dots, R_k$ in $W_I$ such that the following holds: 
    there exists a partition $X_1 \cup \cdots \cup X_{k}$ of $X$ into nonempty sets and preferences $B_j \in L(X_j)$ for $j \in [k]$ such that every $R_i$ for $i \in [k]$ respects the preferences of every $B_j$ (that is, $R_i|_{X_j} = B_j$ for all $j$) and satisfies 
    \[
    X_{i} \succ_{R_i} X_{i+1} \succ_{R_i} \cdots \succ_{R_i} X_{k} \succ_{R_i} X_1 \succ_{R_i} \cdots \succ_{R_i} X_{i-1}.
    \]
    In this case, the preferences $B_j$ for $j \in [k]$ are called the \emph{blocks} of $R_1, \dots, R_k$.
\end{definition}
\begin{definition}[Cyclic Profile]\label{Def:k-cyclic-profile}
    Let $1 \le k \le m$. A \emph{$k$-cyclic profile} with synchrony $\sync$ is a profile $\bfR \in W_I^{\nbar(\sync)}$ such that $W_I$ has a $k$-cyclic preference list $R'_1, \dots, R'_k$ and there is an equitable partition\footnote{A partition $\mathcal{P}$ of a finite set $S$ is \emph{equitable} if $|P| \in \left\{ \floor{|X|/|\mathcal{P}|}, \ceil{|X|/|\mathcal{P}|} \right\}$ for all $P \in \mathcal{P}$. We allow empty sets in this partition since we may have $k > \nbar(\sync)$, forcing at least one of the sets to be empty.} $A_1 \cup \cdots \cup A_k$ of $[\nbar(\sync)]$ where for all $i \in [k]$ and $j \in A_i$, we have $R_j = R'_i$. 
    
    The set of all $k$-cyclic profiles with synchrony $\sync$ is denoted $\calC_k^\sync$. Finally, let 
    \[
    \calC^\sync \triangleq \bigcup_{\frac{\nbar(\sync)}{t} \le k \le m} \calC^\sync_k.
    \]
\end{definition}



Let us show that this definition of cyclic profiles satisfies the intuition stated above.
\begin{lemma}\label{Lem:cyclic-implies-safe}
    If $\bfR \in \calC^\sync$, then for all $i \in [\nbar(\sync)]$, we have $\safe_i(\bfR) = \{R_i\} \cap W_O$.
\end{lemma}
\begin{proof}
    Let $\bfR \in \calC^\sync$ and let $i^* \in [\nbar(\sync)]$. It is obvious from the definition of $\safe_{i^*}(\bfR)$ that $\{R_{i^*}\} \cap W_O \subseteq \safe_{i^*}(\bfR)$. Now suppose $S \in \safe_{i^*}(\bfR)$. Clearly $S \in W_O$, so it remains to show $S = R_{i^*}$. Since $\bfR \in \calC^\sync$, there exists some integer $k$ with $\frac{\nbar(\sync)}{t} \le k \le m$, and there exists a $k$-cyclic preference list $R_1', \dots, R_k'$ along with an equitable partition $A_1 \cup \cdots \cup A_k$ of $[\nbar(\sync)]$ satisfying Definition \ref{Def:k-cyclic-profile}; let $j^* \in [k]$ be the unique integer such that $i^* \in A_{j^*}$. Since $R_1', \dots, R_k'$ is $k$-cyclic, there exists a partition $X_1 \cup \cdots \cup X_k$ of $X$ into nonempty sets, and $B_j \in L(X_j)$ for $j \in [k]$ that satisfies the properties in Definition \ref{Def:k-cyclic-pref-list}. 
    
    Since every $i \in [\nbar(\sync)]$ and $j \in [k]$ satisfy $R_i|_{X_j} = B_j \in L(X)$, it is easy to see that $S|_{X_j} = B_j$, as $S \in \safe_{i^*}(\bfR)$. Let $j_1, j_2 \in [k]$ such that $X_{j_1} \succ_{R_{i^*}} X_{j_2}$ and there exists no $j_3 \in [k]$ such that $X_{j_1} \succ_{R_{i^*}} X_{j_3} \succ_{R_{i^*}} X_{j_2}$. Notice that $R_{i^*} = R'_{j^*}$ since $i^* \in A_{j^*}$. It is easy to show that for all $j \in [k]$, we have $X_{j_1} \succ_{R'_{j}} X_{j_2}$ if and only if $j \ne j_2$; this implies that for all $i \in [\nbar(\sync)]$, we have $X_{j_1} \succ_{R_{i}} X_{j_2}$ if and only if $i \notin A_{j_2}$. In particular, this shows that $i^* \notin A_{j_2}$. Furthermore, since $A_1 \cup \cdots \cup A_k$ is an equitable partition of $[\nbar(\sync)]$ and $k \ge \frac{\nbar(\sync)}{t}$, we have
    \[
    |A_{j_2}| \le \ceil{\frac{\nbar(\sync)}{k}} \le \ceil{\frac{\nbar(\sync)}{\nbar(\sync)/t}} = t.
    \]
    We now let $T \triangleq [\nbar(\sync)] \setminus A_{j_2}$.
    Hence $i^* \in T$ and $|T| \ge \nbar(\sync) - t$. It follows from Definition \ref{Def:safe-area} that $S \in \unanimity(\{R_\alpha: \alpha \in T\})$. By Definition \ref{Def:unanimity-set}, it follows that $X_{j_1} \succ_{S} X_{j_2}$. 

    Since $R'_{j^*} = R_{i^*}$, the definition of $j^*$ and Definition \ref{Def:k-cyclic-pref-list} show that 
    \[
    X_{j^*} \succ_{R_{i^*}} X_{j^*+1} \succ_{R_{i^*}} \cdots \succ_{R_{i^*}} X_{k} \succ_{R_{i^*}} X_1 \succ_{R_{i^*}} \cdots \succ_{R_{i^*}} X_{j^*-1}.
    \]
    Since the choice of $j_1, j_2$ was arbitrary in the above argument, this implies that
    \[
    X_{j^*} \succ_{S} X_{j^*+1} \succ_{S} \cdots \succ_{S} X_{k} \succ_{S} X_1 \succ_{S} \cdots \succ_{S} X_{j^*-1}.
    \]
    Thus, because $S|_{X_j} = B_j = R_{i^*}|_{X_j}$ for all $j \in [k]$, we obtain $S = R_{i^*}$, as desired.
\end{proof}

We are now ready to state and prove our main impossibility theorem.

\begin{theorem}[Main]\label{Thm:main-impossibility}
    Let $k < n$. Then there is no algorithm solving $k$-set preference aggregation in the synchrony model $\sync$ if there exists a $j$-cyclic profile in $\calC^\sync$ for some $j > k$. Similarly, there is no algorithm solving $\epsilon$-approximate preference aggregation in the synchrony model $\sync$ if $\calC^\sync \ne \emptyset$ and $\epsilon < \max_{\bfR \in \calC^\sync} \diam_d(\bfR)$.
\end{theorem}
\begin{proof}
    Suppose $k < \nbar(\sync)$, which is always true if $\sync = \SYNC$.
    Suppose $\bfR \in \calC^\sync$ is a $j$-cyclic profile for some $j > k$.
    Suppose $A$ is an algorithm that solves $k$-set preference aggregation in the $\sync$ synchrony model.
    Then the map $\mathtt{F}_A^\sync$ satisfies $k$-set agreement and $(\nbar(\sync)-t)$-unanimity by Observation \ref{Lem:k-set-reduction} and Lemma \ref{Lem:unanimity-reduction}.
    Since $\bfR$ is $j$-cyclic and $j > k$ and $k < \nbar(\sync)$, there exists a set $S \subseteq [\nbar(\sync)]$ such that $|S| = k+1$ and every $R_i$ is distinct over all $i \in S$. Then by Lemma \ref{Lem:map-unanimity-to-safe-area}, for all $i \in S$, we have $\mathtt{F}_A^\sync(\bfR)_i \in \safe_i(\bfR)$. By Lemma \ref{Lem:cyclic-implies-safe},
    $\safe_i(\bfR) \subseteq \{R_i\}$ for all $i \in S$, which implies
    that $\mathtt{F}_A^\sync(\bfR)_i = R_i$ for all $i \in S$. Since $|S| = k+1$, this shows that $\mathtt{F}_A^\sync$ does not satisfy $k$-set agreement, a contradiction. This proves the result when $k < \nbar(\sync)$.
    
    Now, suppose $k \ge \nbar(\sync)$, so $\sync = \ASYNC$. Suppose there exists a $j$-cyclic profile in $\calC^\sync = \calC^\ASYNC$ for some $k < j \le n$. This profile can be extended to a $j$-cyclic profile $\bfR \in W_I^n$. Since $j > \nbar(\ASYNC) = n-t$, we know $j \ge n-t+1 \ge \frac{n}{t}$, so that $\bfR \in \calC^\SYNC$. Since we have already proved the synchronous case, this shows that there is no algorithm that solves $k$-set preference aggregation in the $\SYNC$ synchrony model, so certainly no algorithm solves the task in the $\ASYNC$ model (such an algorithm necessarily allows synchronous executions).

    For the second part, suppose $\calC^\sync \ne \emptyset$ and $\epsilon < \max_{\bfR \in \calC^\sync} \diam_d(\bfR)$. Suppose $A$ is an algorithm that solves $\epsilon$-approximate preference aggregation in the $\sync$ synchrony model.
    Then the map $\mathtt{F}_A^\sync$ satisfies $\epsilon$-approximate agreement and $(\nbar(\sync)-t)$-unanimity by Observation \ref{Lem:approx-reduction} and Lemma \ref{Lem:unanimity-reduction}.
    Pick any $\bfR \in \argmax_{\bfR' \in \calC^\sync} \diam_d(\bfR')$. Let $i,j \in [\nbar(\sync)]$ such that $d(R_i, R_j) = \diam_d(\bfR)$. Then $d(R_i, R_j) > \epsilon$. By Lemma \ref{Lem:map-unanimity-to-safe-area}, $\mathtt{F}_A^\sync(\bfR)_i \in \safe_i(\bfR)$ and $\mathtt{F}_A^\sync(\bfR)_j \in \safe_j(\bfR)$. Lemma \ref{Lem:cyclic-implies-safe} implies
    $\safe_i(\bfR) \subseteq \{R_i\}$ and $\safe_j(\bfR) \subseteq \{R_j\}$, so that
    that $\mathtt{F}_A^\sync(\bfR)_i = R_i$ and $\mathtt{F}_A^\sync(\bfR)_j = R_j$. It follows that
    \[
    d(\mathtt{F}_A^\sync(\bfR)_i, \mathtt{F}_A^\sync(\bfR)_j) = d(R_i, R_j) > \epsilon,
    \]
    which contradicts the $\epsilon$-agreement property of $\mathtt{F}_A^\sync$. The theorem follows.
\end{proof}

Let us now show some consequences of this theorem on special cases. First, we consider $k$-set preference aggregation on \emph{full domains}, which contain all strict preferences on $X$.
\begin{corollary}[$k$-Set Impossibility: Full Domain]\label{Cor:k-set}
    Suppose $L(X) \subseteq W_I \subseteq P(X)$. If $m \ge \frac{\nbar(\sync)}{t}$ and $k < \min\{m, n\}$, then there is no algorithm solving $k$-set preference aggregation on $W_I, W_O$ in the synchrony model $\sync$.
\end{corollary}
\begin{proof}
    Suppose $m \ge \frac{\nbar(\sync)}{t}$ and $k < \min\{m, n\}$, so $k < m$ and $k < n$. Write $X = \{a_1, \dots, a_m\}$. Consider an $m$-cyclic preference list $R_1, \dots, R_m$ in $L(X)$ given by setting each $X_i = \{a_i\}$ and $B_i$ to be the only preference in $L(X_i)$ and using the relations in Definition \ref{Def:k-cyclic-pref-list}. 
    Let $\bfR \in L(X)^{\nbar(\sync)}$ be an $m$-cyclic profile constructed using Definition \ref{Def:k-cyclic-profile} and any equitable partition $A_1 \cup \cdots \cup A_m$ of $[\nbar(\sync)]$. Since $m \ge \frac{\nbar(\sync)}{t}$, we know that $\bfR \in \calC^{\sync}$. The result follows from Theorem \ref{Thm:main-impossibility} since $\bfR$ is $m$-cyclic and $k < n$ and $k < m$.
\end{proof}

For our analysis of the $\epsilon$-approximate preference aggregation problem, we consider the Kendall tau metric (Definition \ref{Def:kendall-tau}) and Spearman's footrule metric (Definition \ref{Def:spearman-footrule}) on $L(X)$. 

The assumption that $j$ is even in the following result is only for simplicity of algebraic expressions, and is not fundamental to the result itself.
\begin{corollary}[$\epsilon$-Approximate Kendall Tau: General Impossibility]\label{Cor:general-kendall-tau}
    Suppose there exists a $j$-cyclic profile $\bfR \in \calC^\sync$ for some $j \ge \frac{\nbar(\sync)}{t}$ such that each block of $\bfR$ is on at least $\ell \ge 1$ alternatives. If $\epsilon < \floor{j^2/4} \ell^2$, then no algorithm solves $\epsilon$-approximate preference aggregation on $W_I, W_O, \mathtt{KT}$ in the $\sync$ synchrony model. In particular, if $j$ is even and $\delta \triangleq \ell \cdot \frac{j}{m}$, then $\epsilon$-approximate preference aggregation is impossible in the $\sync$ synchrony model for
    \[
    \epsilon < \frac{\delta^2}{2} \cdot \diam_{\mathtt{KT}}(W_I).
    \]
\end{corollary}
\begin{proof}
    Since there exists a $j$-cyclic profile for some $j \ge \frac{\nbar(\sync)}{t}$, then $W_I$ has a $j$-cyclic preference list $R_1', \dots, R_j'$; let $X_1, \dots, X_j$ be the blocks of $R_1', \dots, R_j'$, each of size at least $\ell$. It follows that there exists a $j$-cyclic profile $\bfR \in \calC^\sync$ (constructed with the preference list $R_1', \dots, R_j'$) such that $R_1', R_{\floor{j/2}+1}' \in \set(\bfR)$.
    Observe that for all $a,b \in X$, we have $R_1'|_{\{a,b\}} \ne R_{\floor{j/2}+1}'|_{\{a,b\}}$ if and only if $a \in X_1 \cup \cdots \cup X_{\floor{j/2}}$ and $b \in X_{\floor{j/2}+1} \cup \cdots \cup X_j$ or vice versa (see Figure \ref{Fig:cyclic-preference}); hence $\mathtt{KT}(R_1', R_{\floor{j/2}+1}') = \left(\sum_{i = 1}^{\floor{j/2}} |X_i|\right) \cdot \left(\sum_{i = \floor{j/2}+1}^{j} |X_i|\right)$. 
    
    Let $\epsilon < \floor{j^2/4} \ell^2$.
    Then, since $\ell = \delta \cdot \frac{m}{j}$ and by definition of a $j$-cyclic preference list,
    \begin{align*}
        \diam_{\mathtt{KT}}(\set(\bfR)) &\ge \mathtt{KT}(R_1', R_{\floor{j/2}+1}')
        = \left(\sum_{i = 1}^{\floor{j/2}} |X_i|\right) \cdot \left(\sum_{i = \floor{j/2}+1}^{j} |X_i|\right)\\
        &\ge \floor{j/2}\cdot \ell \cdot (j - \floor{j/2}) \cdot \ell
        = \floor{j/2} \cdot \ceil{j/2} \cdot \ell^2 = \floor{j^2/4} \cdot \ell^2 > \epsilon.
    \end{align*}
    
    The first part then follows from Theorem \ref{Thm:main-impossibility} since $\epsilon <  \max_{\bfR' \in \calC^\sync} \diam_{\mathtt{KT}}(\set(\bfR'))$.
    For the second part, suppose $j$ is even and $\ell = \delta \cdot \frac{m}{j}$ for some real $\delta > 0$. Then, by the first part, $\epsilon$-approximate agreement is impossible if $\epsilon < \floor{j^2/4} \ell^2$. By Proposition \ref{Prop:kendall-spearman-diam},
    \[
    \floor{j^2/4} \ell^2 = \frac{j^2}{4} \cdot \left(\delta \cdot \frac{m}{j}\right)^2 = \frac{\delta^2}{2} \cdot \frac{m^2}{2} \ge \frac{\delta^2}{2} \cdot \diam_{\mathtt{KT}}(L(X)) \ge \frac{\delta^2}{2} \cdot \diam_{\mathtt{KT}}(W_I).
    \]
    Hence $\epsilon < \frac{\delta^2}{2} \cdot \diam_{\mathtt{KT}}(W_I)$ implies $\epsilon < \floor{j^2/4} \ell^2$, proving the result.
\end{proof}

Consider the $\delta$ value in the context of Corollary \ref{Cor:general-kendall-tau} (and in the next corollary). Observe that $\delta$ is at least the ratio of the \emph{smallest} block size to the \emph{average} block size $\frac{m}{j}$. We necessarily have $\delta < 1$, but the closer $\delta$ is to 1, the more the partition $X_1 \cup \cdots \cup X_j$ of $X$ is evenly distributed and closer to being equitable.

\begin{figure}[ht]
    \centering
    \[
    \begin{array}{c|c}
        R'_1 & R'_{\floor{j/2}+1} \\
        \hline
        B_1 & B_{\floor{j/2}+1} \\
        B_2 & B_{\floor{j/2}+2} \\
        \vdots & \vdots \\
        B_{\floor{j/2}} & B_j \\
        \hline
        B_{\floor{j/2}+1} & B_1 \\
        B_{\floor{j/2}+2} & B_2 \\
        \vdots & \vdots \\
        B_j & B_{\floor{j/2}}
    \end{array}
    \]
    \caption{Visualization of $R_1'$ and $R_{\floor{j/2}+1}'$ in the proofs of Corollaries \ref{Cor:general-kendall-tau} and \ref{Cor:general-spearman-footrule}.}
    \label{Fig:cyclic-preference}
\end{figure}

\begin{corollary}[$\epsilon$-Approximate Spearman Footrule: General Impossibility]\label{Cor:general-spearman-footrule}
    Suppose there exists a $j$-cyclic profile $\bfR \in \calC^\sync$ for some $j \ge \frac{\nbar(\sync)}{t}$ such that each block of $\bfR$ is on at least $\ell \ge 1$ alternatives. If $\epsilon < \floor{j^2/2} \ell^2$, then no algorithm solves $\epsilon$-approximate preference aggregation on $W_I, W_O, \mathtt{SF}$ in the $\sync$ synchrony model. In particular, if $j$ is even and $\delta \triangleq \ell \cdot \frac{j}{m}$, then $\epsilon$-approximate preference aggregation is impossible in the $\sync$ synchrony model for
    \[
    \epsilon < \delta^2 \cdot \diam_\mathtt{SF}(W_I).
    \]
\end{corollary}
\begin{proof}
    The proof of this result is almost identical to the proof of Corollary \ref{Cor:general-kendall-tau}. Let $R_1', \dots, R_j'$ and $\bfR \in \calC^\sync$ and $X_1, \dots, X_j$ be defined as in the proof of Corollary \ref{Cor:general-kendall-tau}. Observe that for every $a \in X_1 \cup \cdots \cup X_{\floor{j/2}}$, we have $|\mathrm{rank}_{R_1'}(a) - \mathrm{rank}_{R_{\floor{j/2}}'}(a)| = \sum_{i=\floor{j/2}+1}^j |X_i|$ and for all $a \in X_{\floor{j/2}+1} \cup \cdots \cup X_j$, we have $|\mathrm{rank}_{R_1'}(a) - \mathrm{rank}_{R_{\floor{j/2}}'}(a)| = \sum_{i=1}^{\floor{j/2}} |X_i|$. See Figure \ref{Fig:cyclic-preference} for a visualization of these observations.

    If $\epsilon < \floor{j^2/2} \cdot \ell^2$, then the observations above imply
    \begin{align*}
    \diam_\mathtt{SF}(\set(\bfR)) &\ge \mathtt{SF}(R_1', R_{\floor{j/2}+1}')
    = 2 \cdot \left(\sum_{i = 1}^{\floor{j/2}} |X_i|\right) \cdot \left(\sum_{i = \floor{j/2}+1}^{j} |X_i|\right)\\
    &= 2 \cdot \mathtt{KT}(R_1', R_{\floor{j/2}+1}')
    \ge 2 \cdot \floor{j^2 / 4} \cdot \ell^2 = \floor{j^2 / 2} \cdot \ell^2 > \epsilon.
    \end{align*}
    The result follows by Theorem \ref{Thm:main-impossibility}. The second part of this corollary holds similarly to the proof of Corollary \ref{Cor:general-kendall-tau}. Suppose $j$ is even and write $\ell = \delta \cdot \frac{m}{j}$. The result immediately follows from the first part of this corollary and the inequality
    \[
    \floor{j^2/2} \ell^2 = \frac{j^2}{2} \cdot \left(\delta \cdot \frac{m}{j}\right)^2 = \delta^2 \cdot \frac{m^2}{2} \ge \delta^2 \cdot \diam_\mathtt{SF}(L(X)) \ge \delta^2 \cdot \diam_\mathtt{SF}(W_I),
    \]
    which follows from the equality $\diam_\mathtt{SF}(L(X)) = \floor{\frac{m^2}{2}}$ in Proposition \ref{Prop:kendall-spearman-diam}.
\end{proof}

We conclude this section by analyzing approximate preference aggregation on the full domain consisting of all strict preferences on $X$.

\begin{corollary}[$\epsilon$-Approximate Kendall Tau: Full Domain Impossibility]\label{Cor:full-domain-kendall-tau}
    Assume $W_I = W_O = L(X)$. If $m \ge \frac{\nbar(\sync)}{t}$, then no algorithm solves $\epsilon$-approximate preference aggregation on $W_I, W_O, \mathtt{KT}$ in the $\sync$ synchrony model, if $\epsilon < \floor{m^2/4}$, and in particular, if $\epsilon < \frac{1}{2} \cdot \diam_\mathtt{KT}(L(X))$.
\end{corollary}
\begin{proof}
    Using the construction in the proof of Corollary \ref{Cor:k-set}, there exists a $m$-cyclic profile $\bfR \in \calC^\sync$. Since the blocks of $\bfR$ form an equitable partition of $X$, each block contains exactly one alternative. Since $m \ge \frac{\nbar(\sync)}{t}$, Corollary \ref{Cor:general-kendall-tau} shows that there is no $\epsilon$-approximate preference aggregation algorithm (on $W_I, W_O, \mathtt{KT}$) in the $\sync$ synchrony model for $\epsilon < \floor{m^2/4} \cdot 1^2 = \floor{m^2/4}$. The second part of the result follows from the following inequality:
    $
    \frac{1}{2}\cdot \diam_\mathtt{KT}(L(X)) = \frac{m^2 - m}{4} \le \floor{\frac{m^2}{4}}.
    $
    The first equality holds by Proposition \ref{Prop:kendall-spearman-diam} and the last inequality is obvious if $m \ge 4$, and one may easily verify that it is true when $m \le 3$.
\end{proof}

\begin{corollary}[$\epsilon$-Approximate Spearman Footrule: Full Domain Impossibility]\label{Cor:full-domain-spearman-footrule}
    Assume $W_I = W_O = L(X)$. If $m \ge \frac{\nbar(\sync)}{t}$, then no algorithm solves $\epsilon$-approximate preference aggregation on $W_I, W_O, \mathtt{SF}$ in the $\sync$ synchrony model, if $\epsilon < \floor{m^2/2}$, and in particular, if $\epsilon < \diam_\mathtt{SF}(L(X))$.
\end{corollary}
\begin{proof}
    Using a similar argument as the previous corollary and by Corollary \ref{Cor:general-spearman-footrule}, we have the following. If $m \ge \frac{\nbar(\sync)}{t}$, then there is no $\epsilon$-approximate preference aggregation algorithm (on $W_I, W_O, \mathtt{SF}$) in the $\sync$ synchrony model for $\epsilon < \floor{m^2/2}$. The second part of the result follows from the equality $\diam_\mathtt{SF}(L(X)) = \floor{m^2 / 2}$, which holds by Proposition \ref{Prop:kendall-spearman-diam}.
\end{proof}

\section{Related Work}
The Arrovian framework in voting theory has already received a lot of attention; however, its intersection with distributed computing appears to be recent.
To the best of our knowledge, distributed combinatorial topology techniques such as the \emph{index lemma} on simplicial complexes were first used in 2022 \cite{sergio2022} to prove the base case ($m=3$, $n=2$) of Arrow's theorem topologically, and then extended via induction. This work was later improved in 2024 \cite{sergio2024} by proving a domain-generalization of Arrow's theorem using only distributed combinatorial topology on \emph{all} cases. Other work has studied social welfare and social choice distributed algorithms satisfying consensus and significantly weaker validity conditions (e.g. preserving unanimity of only the highest ranked alternative) than our unanimity condition \cite{democratic-elections}.

The theory in the Arrovian framework itself is full of rich results. In particular, Arrow's theorem and domain-generalizations thereof have been proved via many combinatorial methods \cite{campbell-chapter1, three-brief-proofs}, as well as techniques from algebraic topology \cite{algebraic-topology-arrow-proof} and fixed-point methods in metric spaces \cite{fixed-point-proof1, fixed-point-proof2}. More generally, (algebraic) topological techniques have proven successful in voting theory. For example, the Gibbard--Satterthwaite theorem---which shows the non-existence of a \emph{social choice function} (inputs being preferences and output being a single alternative) satisfying surjectivity and strategy-proofness---has been proven using combinatorial methods and algebraic topological methods \cite{Gibbard_1998, gibbard-topological-proof}.

On the distributed computing side, the $k$-set agreement task (in which processes must decide on at most $k$ values with a notion of decision validity) was introduced in \cite{chaudhuri-k-set-init}. Other formulations of this task may be found in \cite{k-set-consensus-async}. The $\epsilon$-approximate agreement distributed task on the real line was well-studied in \cite{approx-consensus, approx-consensus3} and was proven to be solvable in fully asynchronous systems with relatively high resilience, despite the consensus unsolvability in such systems with only one crash fault \cite{flp-theorem}. These results were later generalized to \emph{multidimensional} approximate agreement in \cite{mendes-approx-agreement-DISC, mendes-approx-agreement-journal}. Multidimensional approximate agreement has been studied extensively since then \cite{approx-agree-asymp-optimal, approx-agree-async-fallback, approx-agree-async-quadratic, approx-agree-fast, approx-agree-step-complexity}. Discrete versions of approximate agreement have been formulated on graphs and simplicial complexes \cite{byzantine-approx-agree-graphs, wait-free-approx-agree-graphs, approx-agreement-simplicial-complexes}; these tasks are somewhat similar to the $\epsilon$-approximate preference aggregation task discussed in this paper.


\section{Conclusion}
In this paper, we propose two novel distributed tasks in the Arrovian framework and prove strong impossibility results on these tasks in crash-prone, synchronous and asynchronous systems. We adapt previously well-studied distributed tasks---namely, set and approximate agreement---to the context of preference aggregation, by replacing the respective validity properties with unanimity on the correct processes. Our impossibility results are very general, and apply to $k$-set preference aggregation on a full domain, and $\epsilon$-approximate preference aggregation with both the Kendall tau and Spearman footrule metrics.

Using the Kendall tau metric on a domain of strict preferences illuminates a particularly fascinating connection to a more general ``metric space approximate agreement'' task. One may embed the given domain into a higher dimensional Euclidean space by examining the order of each pair of alternatives and mapping it to a binary real. With such an embedding, one may think of the approximate preference aggregation task (with the Kendall tau metric) as a more traditional multidimensional approximate agreement problem, where the convexity and agreement properties are with respect to the Euclidean $L^1$ ``taxicab'' metric (instead of the usual $L^2$ metric) and the definition of convexity is adjusted to a more ``total'' convexity. Many of the lemmas in this paper easily generalize to this metric space framework. Exploration of this more general approximate agreement task would be interesting future work.

Another interesting direction for future work is the following. In our execution map of an asynchronous algorithm, we make a rather arbitrary choice for the set of silent (initially crashed) processes (see the sentence before Definition \ref{Def:alg-reduction-ASYNC}). Using a more rich view and including all possible sets of silent processes in the map (or other techniques) could potentially generate a simplicial complex that captures this information. It would be highly insightful to determine if topological methods could be used in this way to obtain stronger asynchronous results in our Arrovian framework, but also in the more general metric space framework discussed above.

\bibliographystyle{plainurl}
\bibliography{references}

\end{document}